\documentclass[twocolumn,aps,prd,preprintnumbers,superscriptaddress,nofootinbib,10pt]{revtex4-1}
\usepackage{epsfig}
\usepackage{graphicx}
\usepackage{psfrag}
\usepackage{amsmath,amssymb}
\usepackage{colordvi}
\usepackage{amsfonts}
\usepackage{enumerate}
\usepackage{slashed}
\usepackage{color}
\usepackage{xcolor}

\begin{document}
\title{Near Threshold Heavy Quarkonium Photoproduction at Large Momentum Transfer }

\author{Peng Sun}
\affiliation{Department of Physics and Institute of Theoretical Physics,
Nanjing Normal University, Nanjing, Jiangsu 210023, China}

\author{Xuan-Bo Tong}
\affiliation{School of Science and Engineering, The Chinese University of Hong Kong, Shenzhen, Shenzhen, Guangdong, 518172, P.R. China}
\affiliation{University of Science and Technology of China, Hefei, Anhui, 230026, P.R.China}

\author{Feng Yuan}
\affiliation{Nuclear Science Division, Lawrence Berkeley National
Laboratory, Berkeley, CA 94720, USA}

\begin{abstract}
Perturbative QCD is applied to investigate the near threshold heavy quarkonium photoproduction at large momentum transfer. We take into account the contributions from the leading three-quark Fock states of the nucleon. The dominant contribution comes from the three-quark Fock state with one unit quark orbital angular momentum (OAM) whereas that from zero quark OAM is suppressed at the threshold. From our analysis, we also show that there is no direct connection between the near threshold heavy quarkonium photoproduction and the gluonic gravitational form factors of the nucleon. Based on the comparison between our result and recent GlueX data of $J/\psi$ photoproduction, we make predictions for $\psi'$ and $\Upsilon$ (1S,2S) states which can be tested in future experiments. \end{abstract}
\maketitle

\section{Introduction}

Exclusive heavy quarkonium production in high energy photon-proton scattering, \begin{equation}
    \gamma^{(*)} + N\to V +N' \ ,
\end{equation} 
where the incoming photon can be real or virtual, has attracted great attention in hadron physics community. This process is dominated by the two gluon exchange~\cite{Ryskin:1992ui,Brodsky:1994kf} and can be formulated in the generalized parton distribution (GPD)~\cite{Ji:1996ek,Ji:1996nm} framework~\cite{Collins:1996fb,Hoodbhoy:1996zg,Koempel:2011rc,Cui:2018jha,Ivanov:2004vd}. The theory advance has also pushed the perturbative QCD computation of these processes to the next-to-leading order~\cite{Ivanov:2004vd,Chen:2019uit, Flett:2021ghh}.

Recently, there has been a strong interest of this process at the lower end of the energy range near the threshold~\cite{Kharzeev:1995ij,Kharzeev:1998bz,Gryniuk:2016mpk,Hatta:2018ina,Ali:2019lzf,Hatta:2019lxo,Boussarie:2020vmu,Mamo:2019mka,Gryniuk:2020mlh,Wang:2019mza,Zeng:2020coc,Du:2020bqj,Kharzeev:2021qkd,Wang:2021dis,Hatta:2021can,Mamo:2021krl,Kou:2021bez,Guo:2021ibg,Mamo:2021tzd}. In particular, it was argued in Refs.~\cite{Kharzeev:1995ij,Kharzeev:1998bz} that this process can provide a direct access to the so-called trace anomaly contribution to the proton mass, while the origin of the proton mass is of fundamental in QCD strong interaction theory~\cite{Shifman:1978zn,Ji:1994av,Ji:1995sv,Hatta:2018sqd,Metz:2020vxd,Hatta:2020iin,Ji:2021pys,Ji:2021mtz,Lorce:2021xku}. 

In experiments, $J/\psi$ photo-production from the nuclear targets near the threshold have been investigated before~\cite{Gittelman:1975ix,Camerini:1975cy}. More recently, high precision measurements have been carried out by the GlueX collaboration at Jefferson Lab~\cite{Ali:2019lzf}. Future experiments will explore both $J/\psi$ and $\Upsilon$ near threshold photo-production in great details~\cite{Joosten:2018gyo}, including JLab-12GeV~\cite{Dudek:2012vr,Chen:2014psa} and electron-ion colliders (EIC)~\cite{Accardi:2012qut,AbdulKhalek:2021gbh,Anderle:2021wcy}. 

In this paper, we will focus on one of the key aspects of the threshold kinematics that the momentum transfer is relatively large: $-t\sim 2{\rm GeV}^2$ and $10{\rm GeV}^2$ for $J/\psi$ and $\Upsilon$, respectively. Here, $t$ is the momentum transfer squared from the nucleon target. Because of this large momentum transfer, we can apply the QCD factorization argument to compute the scattering amplitude. This factorization follows that of the hadron form factor calculations in perturbative QCD  ~\cite{Lepage:1979za,Brodsky:1981kj,Efremov:1979qk,Chernyak:1977as,Chernyak:1980dj,Chernyak:1983ej,Belitsky:2002kj,Tong:2021ctu}. For the heavy quarkonium production in the final state, the non-relativistic QCD (NRQCD)~\cite{Bodwin:1994jh} will be adopted and the associated color-singlet matrix element of the quarkonium state is responsible for its production in the exclusive process. 

In the perturbative calculations, the quark/gluon propagators in the scattering amplitudes lead to the power behavior for the differential cross section at large momentum transfer~\cite{Brodsky:1973kr,Matveev:1973ra,Ji:2003yj}, which has been commonly assumed in the phenomenological studies of near threshold heavy quarkonium production, see, e.g., Refs.~\cite{Frankfurt:2002ka,Ali:2019lzf,Kharzeev:2021qkd,Wang:2021dis}. In the following, we will provide an explicit calculation to demonstrate this power behavior. 

The hard exclusive processes at large momentum transfer depend on the non-perturbative distribution amplitudes~\cite{Lepage:1979za}. In our derivations, we take into account the contributions from the leading-twist and higher-twist terms of the nucleon distribution amplitudes~\cite{Braun:1999te,Braun:2000kw}. They correspond to the three-quark Fock state light-cone wave functions of the nucleon with zero orbital angular momentum (OAM) and one unit OAM components~\cite{Ji:2002xn}, respectively. Their contributions lead to different power behaviors at large $(-t)$, similar to the nucleon's form factors~\cite{Belitsky:2002kj,Tong:2021ctu}.

We will also take the heavy quark mass limit and apply the following hierarchy in scales: 
\begin{equation}
    W_{\gamma p}^2\sim M_V^2\gg (-t)\gg\Lambda_{QCD}^2\ ,
\end{equation}
where $\Lambda_{QCD}$ for the non-perturbative scale. In addition, throughout the following analysis, we take the threshold limit, i.e., $W_{\gamma p}\sim M_V+M_p$, where $W_{\gamma p}$ represents the center of mass energy and $M_V$ and $M_p$ for the heavy quarkonium and proton masses, respectively. To determine the leading contribution, we introduce a parameter~\cite{Brodsky:2000zc}, 
\begin{equation}
\chi=\frac{M_{V}^2+2M_pM_{V}}{W_{\gamma p}^2-M_p^2} \ ,
\end{equation}
which goes to 1 at the threshold. We will expand the amplitude in terms of $(1-\chi)$. By applying this expansion, we find that the commonly used $1/(-t)^4$ power term for the differential cross section is suppressed by $(1-\chi)$. The dominant contribution at the threshold actually comes from the higher-twist term with $1/(-t)^5$ power behavior. 

As mentioned above, the exclusive heavy quarkonium production has been extensively studied in the GPD framework and the scattering amplitude can be written in terms of the gluon GPDs. In Refs.~\cite{Boussarie:2020vmu,Hatta:2021can,Guo:2021ibg}, the GPD formalism was applied in the the threshold kinematics, where the connection to the gluonic gravitational form factors was explored. One of the major objectives of this paper is to check the connection between the near threshold heavy quarkonium photo-production and the gluonic gravitational form factors. To do that, we compare the differential cross section derived in this paper and those of the gluonic gravitational form factors of the nucleon at large momentum transfer in Ref.~\cite{Tong:2021ctu}. We will show explicitly that there is no direct connection between them. Therefore, approximations have to be made to link the GPD formalism of this process to the gluonic gravitational form factors~\cite{Boussarie:2020vmu,Hatta:2021can,Guo:2021ibg}. A brief summary of our results has been published in Ref.~\cite{Sun:2021gmi}. In the following, we provide more detailed derivations. 

The rest of the paper is organized as follows. In Sec.~II, we will examine the threshold kinematics and apply the expansion method mentioned above to simplify the derivation. In Sec.~III, we take the example of photon scattering off a pion target. The leading Fock state of the pion contains quark and antiquark and the derivation is much simpler compared to the nucleon case. Sec.~IV and V will be dedicated to the nucleon case. In Sec.~IV, we study the contribution from the leading component of the nucleon distribution amplitude and show that its contribution is actually suppressed in the threshold limit. In Sec.~V, we perform the analysis of higher-twist component of the nucleon distribution amplitude and show that its contribution to the differential cross section does not vanish at the threshold. In Sec.~VI, we discuss the interpretation and consequence of our derivations. We conclude that there is no direct connection between the near threshold photo-production of heavy quarkonium and the gluonic gravitational form factors of the nucleon. In Sec.~VII, we provide phenomenological applications of our derivations. Predictions on $\psi'$ and $\Upsilon$ will be presented for future experiments based on the comparison between our results and the GlueX data on near threshold $J/\psi$ production at JLab. Finally, we summarize our paper in Sec.~VIII.

\section{Near Threshold Kinematics}

\begin{figure}[tpb]
\includegraphics[width=0.7\columnwidth]{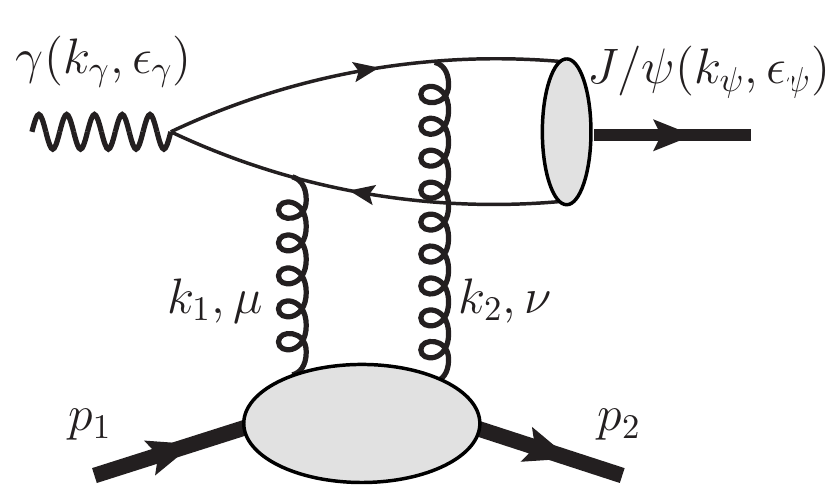}
    \caption{Schematics of two-gluon exchange contribution to the threshold heavy quarkonium production.}
    \label{fig:threshold}
\end{figure}

The typical Feynman diagram of the two-gluon exchange contributions to the near threshold heavy quarkonium photoproduction is shown in Fig.~\ref{fig:threshold},
\begin{equation}
    \gamma (k_\gamma)+N(p_1)\to J/\psi (k_\psi)+N'(p_2) \ ,
\end{equation}
where we have used $J/\psi$ as an example. In order to make the near threshold expansion more evident, it is useful to examine the relevant kinematics for the scattering amplitude. The center of mass energy and momentum transfer squared can be written as,
\begin{eqnarray}
&&W_{\gamma p}^2=(k_\gamma+p_1)^2=(k_\psi+p_2)^2\sim M_V^2 \ ,\\
&&|t|=|(p_2-p_1)^2|\ll M_V^2 \ .
\end{eqnarray}
Therefore, we will have the following approximations around the threshold kinematics,
\begin{eqnarray}
&&p_1\cdot k_\gamma\sim p_1\cdot k_\psi\sim M_V^2\ ,\\
&&p_2\cdot k_\gamma\sim p_2\cdot k_\psi\ll M_V^2\ .
\end{eqnarray}
In addition, applying the heavy quark mass limit of $M_V^2\gg (-t)$, we find that the invariant mass of the $t$-channel two gluons is much smaller than heavy quarkonium mass. We will also take the approximation of $M_c\approx M_V/2$ in the non-relativistic limit of the heavy quarkonium system.

The quark propagators in the upper part of the Feynman diagram of Fig.~\ref{fig:threshold} are all in order of $1/M_V$. For example, one of the quark propagators can be simplified as
\begin{eqnarray}
\frac{1}{\left(k_1-k_\psi/2\right)^2-M_c^2}&=&\frac{1}{-k_1\cdot k_\gamma -k_1\cdot k_2}\nonumber\\
&\approx&\frac{1}{-k_1\cdot k_\gamma}\ ,
\end{eqnarray}
where we have applied $|k_1^2|\sim |k_2^2|\sim |k_1\cdot k_2|\sim |t|\ll M_V^2$. Because $k_1$ carries certain momentum fraction of the incoming nucleon, $k_1\cdot k_\gamma$ will be order of $M_V^2$. Similarly, we have
\begin{eqnarray}
\frac{1}{\left(k_2-k_\psi/2\right)^2-M_c^2}
&\approx&\frac{1}{-k_2\cdot k_\gamma}\ .
\end{eqnarray}
The following propagator will also show up in some of the Feynman diagrams,
\begin{eqnarray}
    \frac{1}{\left(k-k_\psi/2\right)^2-M_c^2}=\frac{1}{-k\cdot k_\gamma}\approx \frac{2}{-M_V^2}\ ,
\end{eqnarray}
where $k=k_1+k_2=p_1-p_2$. In the center of mass frame, $k$ is dominated by $p_1$ because $p_2$ is soft. 

Applying the above approximations, we can simplify the photon-heavy quarkonium transition amplitude. Let us define $\mu$ and $\nu$ for the polarization indices for $k_1$ and $k_2$, respectively, and $\epsilon_\gamma$ and $\epsilon_\psi$ for the photon polarization and $J/\psi$ polarization vectors, respectively. To further simplify the derivation, we choose the physical polarization for the incoming photon,
\begin{equation}
    \epsilon_\gamma\cdot k_\gamma=0,~~\epsilon_\gamma\cdot p_1=0 \ .
\end{equation}
With this choice, we notice that the contributions from $\epsilon_\gamma\cdot k_1$ and $\epsilon_\gamma\cdot k_2$ are also suppressed in the heavy quark mass limit. Therefore, we will drop these terms as well. We emphasize, all these approximations have been cross checked by a full computation.

Finally, we have the following expression for the amplitude from the heavy quarkonium side,
\begin{equation}
{\cal M}_{\psi,ab}^{\mu\nu}=\frac{\delta^{ab}N_\psi\left[\epsilon_\psi^*\cdot \epsilon_\gamma {\cal W}_T^{\mu\nu}+\epsilon_\psi^*\cdot k {\cal W}_L^{\mu\nu}+{\cal W}_S^{\mu\nu}\right]}{k_1\cdot k_\gamma k_2\cdot k_\gamma} \ ,
\end{equation}
where $a$ and $b$ represent the color indices for the $t$-channel gluons. 
In the above equation, $N_\psi$ is defined as
\begin{equation}
    N_\psi=-\frac{4e_c e g_s^2}{\sqrt{ N_c M_V^3}} \psi_{J}(0)\ ,
    \label{eq:Npsi}
\end{equation}
where $\psi_J(0)$ is the wave function of $J/\psi$ at the origin and is related to the NRQCD matrix element~\cite{Bodwin:1994jh}. The tensor structures ${\cal W}_{T,L,S}^{\mu\nu}$ are defined as
\begin{eqnarray}
{\cal W}_T^{\mu\nu}&=&
-k_1\cdot k_\gamma k_2\cdot k_\gamma g^{\mu\nu}-k_1\cdot k_2 k_\gamma^\mu k_\gamma^\nu\nonumber\\
&&+k_1\cdot k_\gamma k_2^\mu k_\gamma^\nu+k_2\cdot k_\gamma k_1^\nu k_\gamma^\mu\nonumber\\
{\cal W}_L^{\mu\nu}&=&k_1\cdot k_\gamma \epsilon_\gamma^\nu k_2^\mu+k_2\cdot k_\gamma \epsilon_\gamma^\mu k_1^\nu \nonumber\\
{\cal W}_S^{\mu\nu}&=&-k_1\cdot k_2\left(k_1\cdot k_\gamma \epsilon_\psi^{*\mu} \epsilon_\gamma^\nu+k_2\cdot k_\gamma \epsilon_\psi^{*\nu} \epsilon_\gamma^\mu\right.\nonumber\\
&&~~\left.+k_1\cdot \epsilon_\psi^* k_\gamma^\nu \epsilon_\gamma^\mu+k_2\cdot \epsilon_\psi^* k_\gamma^\mu \epsilon_\gamma^\nu
\right)\ .
\end{eqnarray}
where ${\cal W}_T$ and ${\cal W}_L$ represent the amplitudes for a transversely polarized and longitudinal polarized heavy quarkonium in the final state, respectively, whereas ${\cal W}_S$ for a subleading term. 

Clearly the above amplitude is symmetric under $k_1,\mu\leftrightarrow k_2, \nu$. In the above equation, the first term is the leading contribution in the heavy quark mass limit at the threshold. The second and third terms are subleading contributions. 

We have also carried out an important cross check for the above results. We compute the full amplitude without any approximation. We then take the leading contribution of the differential cross section (the amplitude squared) in the heavy quark mass limit and threshold limit, and obtain the same result. 
\subsection{Vanishing of Three-gluon Exchange Contribution}

\begin{figure}[tpb]
\includegraphics[width=0.7\columnwidth]{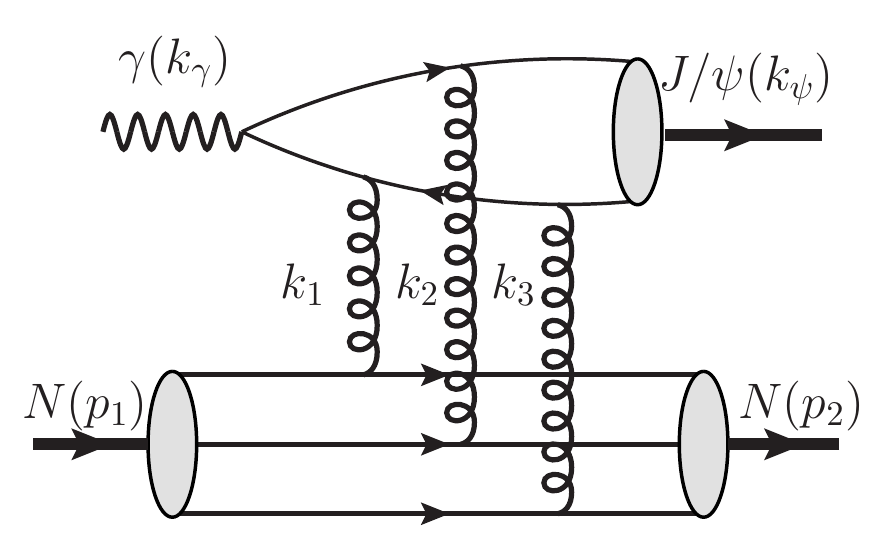}
    \caption{Typical Feynman diagram from three-gluon exchange. These diagrams vanish because of the $C$-parity conservation.}
    \label{fig:threegluon}
\end{figure}

Before we start our derivations of the threshold scattering amplitudes, we would like to comment on the three-gluon exchange contributions. The two-gluon and three-gluon exchange diagrams were considered in Ref.~\cite{Brodsky:2000zc} for the threshold production of $J/\psi$ and it was argued that the three-gluon exchange diagrams dominate the differential cross section contributions. 

However, we find that the three-gluon exchange diagrams do not contribute in our framework, due to the $C$-parity conservation. This is because the three gluons from the nucleon side carry symmetric color structure (such as $d_{abc}$) while those from the heavy quarkonium ($J/\psi$) side are antisymmetric (such as $f_{abc}$), where $a$, $b$ and $c$ represent the color indices for the three gluons in the $t$-channel, respectively. Explicitly, from the nucleon side, we have, as shown in Fig.~\ref{fig:threegluon},
\begin{eqnarray}
\epsilon^{ijk}\epsilon^{lmn}T^a_{il}T^b_{jm}T^c_{kn}\propto d^{abc} \ ,
\end{eqnarray}
where $ijk$ and $lmn$ represent the color indices for the initial and final three quarks, respectively. Here, we have applied the anti-symmetric color structure for the three-quark Fock state wave function of the nucleon~\cite{Ji:2002xn}. On the other hand, for the heavy quarkonium side, we have, instead
\begin{eqnarray}
{\rm Tr}\left[T^aT^bT^c\right]=\frac{1}{4}\left(d^{abc}+if^{abc}\right) \ .
\end{eqnarray}
However, because of $J/\psi$ is in the $1^{--}$ state, the photon-$J/\psi$ transition amplitude vanishes for the symmetric color configuration with three gluons, i.e., $d^{abc}$ term from the above vanishes. Combining this with the color structure from the nucleon side, we conclude the three-gluon exchange diagrams do not contribute.

\section{Pion Case}

\begin{figure}[tpb]
\includegraphics[width=0.7\columnwidth]{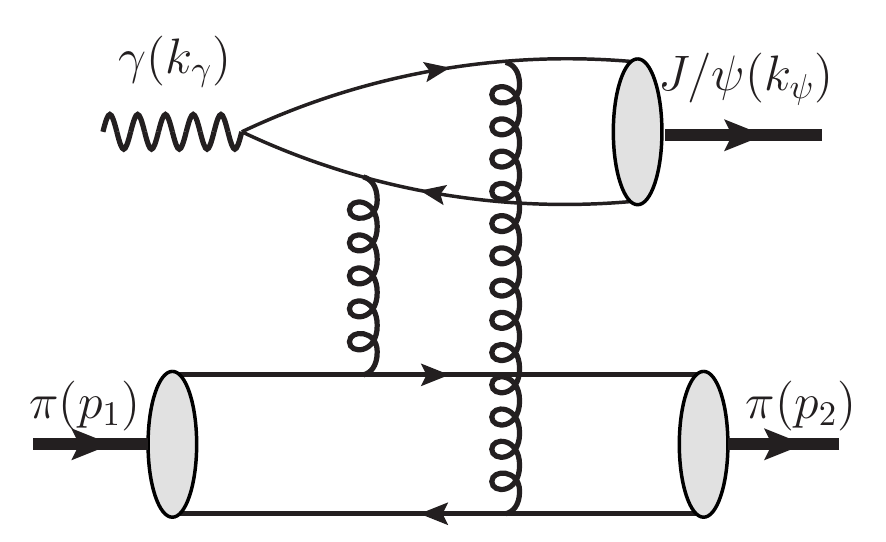}
    \caption{The Feynman diagram contribution to the exclusive $\gamma \pi\to \pi J/\psi$ at large momentum transfer. The two gluons attach to the quark and antiquark lines, respectively.}
    \label{largetdiagrampion}
\end{figure}

In this section, we take the example of pion case to show the detailed of our derivations.  In this case, we have photon scatters on the pion target and produces a $J/\psi$ in the final state close to the threshold,
\begin{equation}
    \gamma +\pi\to J/\psi +\pi \ ,
\end{equation}
where the dominant contribution is again a two-gluon exchange diagram. The two gluons attach to the two quark lines from the pion target, as shown in Fig.~\ref{largetdiagrampion}.

Considering the leading Fock component of the pion, we have  
\begin{eqnarray}
  |\pi^+\rangle_{u\overline d} &=& \int d[1]d[2] \psi_{u\overline d}(1,2)  
\frac{\delta_{ij}}{\sqrt{3}}\left[u^\dagger_{\uparrow i}(1)
  \overline d^\dagger_{\downarrow j}(2) \right.\nonumber\\
  &&\left. - u^\dagger_{\downarrow i}(1)
\overline d^\dagger_{\uparrow j}(2)\right]
 |0\rangle \,  
\end{eqnarray}
where $i$ and $j=1,2,3$ are the color indices, and $\uparrow$ and $\downarrow$ label quark light-cone helicity $+1/2$ and $-1/2$, respectively. The color factor $\delta_{ij}/\sqrt{3}$ is normalized to 1. The light-cone wave function amplitude $\psi_{u\bar d}(1,2)$ is a function of quark momenta with argument 1 representing $x_1$ and $q_{1\perp}$ and so on. Since the momentum conservation implies $\vec{q}_{1\perp}+\vec{q}_{2\perp}=0$ and $x_1+x_2=1$, $\psi_{u\bar d}(1,2)$ depends on variables $x_1$ and $q_{1\perp}$ only. The integration in the above equation is defined as,
\begin{eqnarray}
    \int d[1]d[2]=\int \frac{d^2q_{1\perp}}{(2\pi)^3}\frac{dx_1}{2\sqrt{x_1(1-x_1)}} \ .
\end{eqnarray}
From the light-cone wave function, we obtain the distribution amplitude,
\begin{eqnarray}
    \phi(x)=\int \frac{d^2 q_{1\perp}}{(2\pi)^3}  \psi_{u\bar d}(1,2)\ .
\end{eqnarray}
The final scattering amplitude of $\gamma+\pi^+\to J/\psi+\pi^+$ can be computed in terms of the above distribution amplitude of pion,
\begin{align}
    {\cal A}^\pi=  &  \int d x_1 d y_1 \phi^*(y_1)\phi(x_1)  {\cal M}_\psi^{\mu\nu}(\epsilon_\gamma,\epsilon_{\psi},x_1,y_1)
    \nonumber \\
    &\times \frac{-g_s^2C_F}{2k_1^2k_2^2}{\rm Tr}\left[\slashed{p}_2\gamma^\mu\slashed{p}_1\gamma^\nu\right]
     \ ,
\end{align}
where ${\cal M}_\psi^{\mu\nu}$ has been given in the previous section. 

\subsection{Threshold Expansion}

At the threshold, the amplitude squared can be further simplified as 
\begin{eqnarray}
    |\overline{{\cal A}^\pi}|^2= G_\psi G_{\pi}(t) G_{\pi}^*(t) \ ,\label{eq:pion}
\end{eqnarray}
where the spin sum and average have been applied. Here, $G_\psi$ is defined as 
\begin{equation}
G_\psi=|
N_\psi   |^2=\frac{384 \pi^2 e_c^2 \alpha(4\pi \alpha_s)^2}{ N_c^2 M_\psi^3} \langle  0 | {\cal O}^{\psi}({}^3S_1^{(1)}) |0\rangle \ , \label{eq:gpsi}
\end{equation} 
where $\langle 0| {\cal O}({}^3S_1) |0\rangle$ is the color-singlet NRQCD matrix element for $J/\psi$. $G_\pi(t)$ is defined as  
\begin{equation}
    G_\pi(t)=\frac{8\pi\alpha_sC_F}{t}\int dx_1 dy_1\phi^*(y_1)\phi(x_1)\frac{1}{x_1\bar x_1 y_1\bar y_1} \ ,\label{eq:gpi}
\end{equation}
where $C_F=(N_c^2-1)/2N_c$, $\bar x_1=1-x_1$ and $\bar y_1=1-y_1$. Here, we have neglected high order corrections of $t_\psi=-t/M_V^2$.

\subsection{Compared to the Gravitational Form Factors}

We now compare the above result to the gluonic gravitational form factors at large momentum transfer, which have been computed in Ref.~\cite{Tong:2021ctu}. For convenience, we list the results below. The gluonic gravitation form factors of the pion are defined as 
\begin{align}
&\langle p_2 | T_g^{\mu\nu}|p_1\rangle
  =2 \bar  P^\mu  \bar  P^\nu A_g^{\pi}(t)+\frac{1}{2} ( \Delta^\mu \Delta^\nu-g^{\mu\nu}\Delta^2) C_g^{\pi}(t) \nonumber\\
  &+2m^2 g^{\mu\nu }\overline{C}_g^\pi(t)\ , \label{EMT:pion}
 \end{align} 
where $T_g^{\mu\nu}$ is the gluonic energy-momentum tensor in QCD and $m$ represents the pion mass.
Here, $\bar P=(p_1+p_2)/2$ is the average momentum, $\Delta=p_2-p_1$ is the momentum transfer and hence $t=\Delta^2$. From the results of Ref.~\cite{Tong:2021ctu}, we find
\begin{align}
&A^\pi_g(t)=C^\pi_g(t)=\frac{4m^2}{t}\overline{C}_g^\pi(t)\label{eq:pionformfactor}\\
&=\frac{4\pi \alpha_s C_F}{-t}\int dx_1 dy_1\phi^*(y_1)\phi(x_1) \left(\frac{1}{x_1 \bar x_1}+\frac{1}{y_1 \bar y_1} \right) \ . \nonumber
\end{align}

From the above results, we find that there is no direct connection between $G_\pi(t)$ of Eq.~(\ref{eq:gpi}) and any of the gravitational form factors of $A_g^\pi(t)$, $C_g^\pi(t)$ or $\overline{C}_g^\pi(t)$ (Eq.~(\ref{eq:pionformfactor})). This indicates that we can not directly interpret the near threshold heavy quarkonium photo-production in terms of the gluonic gravitational form factors. 

\subsection{Compared to the GPD Formalism}

As mentioned in the Introduction, the photo-production of heavy quarkonium has been derived in the GPD framework. If we extend these derivations to the near threshold kinematics, we obtain
\begin{eqnarray}
    {\cal A}^\pi&=& N_\psi\epsilon_\psi^*\cdot \epsilon_\gamma \int_{-1}^{1} dx \frac{H_g^\pi(x,\xi,t)}{(x+\xi-i \varepsilon)(x-\xi+i \varepsilon)} \ ,
\end{eqnarray}
for the pion target, where $N_\psi$ has been given in Eq. (\ref{eq:Npsi}) and $\xi$ is the skewness parameter. In the threshold limit we take $\xi=1$. In the above equation, $H_g^\pi$ represents the GPD gluon distribution of the pion. The GPD gluon distribution at large momentum transfer can be calculated in terms of the distribution amplitudes as that of the quark GPD in Ref.~\cite{Hoodbhoy:2003uu}, for which we list in Appendix A. If we substitute the result of $H_g^\pi(x,\xi,t)$ from there, we will be able to reproduce the scattering amplitude result from the direct computation in the above subsection A. This provides a useful cross check for our derivations.

\section{Nucleon Case: twist-three contributions}

\begin{figure}[tpb]
\includegraphics[width=0.7\columnwidth]{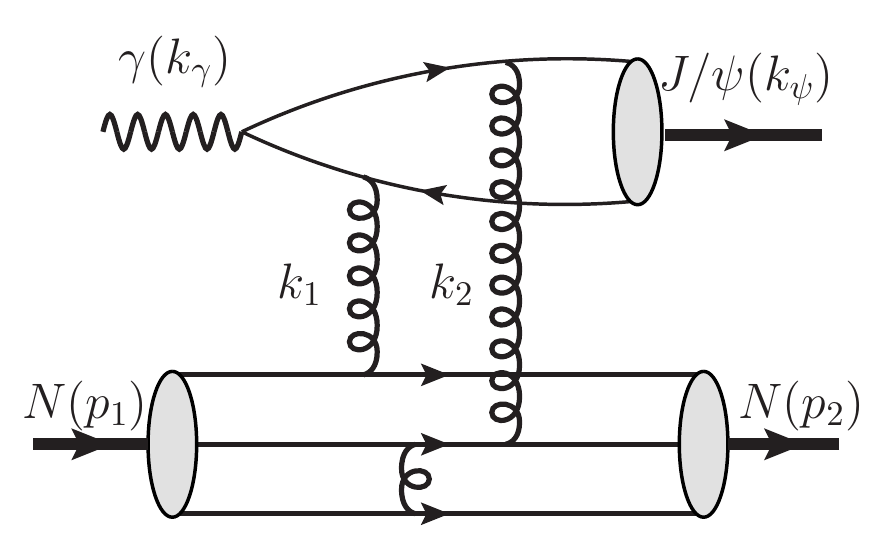}
    \caption{Typical Feynman diagram contributions to the threshold $J/\psi$ photoproduction at large momentum transfer from two-gluon exchange.}
    \label{fig:twogluon}
\end{figure}

Now we turn to the proton cases. We show the typical Feynman diagram in Fig.~\ref{fig:twogluon}. To compute these diagrams, we follow the factorization argument for the hard exclusive processes~\cite{Brodsky:1981kj}, where the leading contributions come from the three quark Fock state of the nucleon. The three-quark states can be further classified into zero orbital angular momentum (OAM) and nonzero OAM components~\cite{Ji:2002xn}.  We will first examine the contribution from zero OAM component. This corresponds to the twist-three contribution from the nucleon's distribution amplitude.

\subsection{Three-quark Fock State with Zero OAM}

Because there is no quark OAM, the total quark spin equals to the nucleon spin. The associated light-cone wave function amplitude is defined as
\begin{eqnarray}
&&  |P\uparrow\rangle_{1/2} = \int d[1]d[2]d[3]\left(\tilde \psi^{(1)}(1,2,3)
         \right) \nonumber \\
         &&  \times  \frac{\epsilon^{ijk}}{\sqrt{6}} u^{\dagger}_{i\uparrow}(1)
             \left(u^{\dagger}_{j\downarrow}(2)d^{\dagger}_{k\uparrow}(3)
            -d^{\dagger}_{j\downarrow}(2)u^{\dagger}_{k\uparrow}(3)\right)
         |0\rangle \ , 
         \label{eq:WFzeroOAM}
\end{eqnarray}
where $ijk$ represent the color indices for the three quarks, respectively, and the measure for the quark momentum is,
\begin{eqnarray}
&&    d[1]d[2]d[3] = \sqrt{2}\frac{dx_1dx_2dx_3}{\sqrt{2x_1 2x_2 2x_3}}
                  \frac{d^2\vec{q}_{1\perp}d^2
             \vec{q}_{2\perp}d^2\vec{q}_{3\perp}}{(2\pi)^9} 
       \nonumber \\
                  && \times (2\pi)^3\delta(1-x_1-x_2-x_3)\delta^{(2)}
                   (\vec{q}_{1\perp}+\vec{q}_{2\perp}+\vec{q}_{3\perp}) \ .
\end{eqnarray}
By integrating over the transverse momenta $q_{i\perp}$, we obtain the twist-three distribution amplitude~\cite{Braun:1999te}
\begin{eqnarray}
    \Phi_3(x_1,x_2,x_3) &=& -2\sqrt{6}\int  [dq_\perp]\tilde \psi^{(1)}(1,2,3) \ ,
\end{eqnarray}
where $[dq_\perp]=\frac{d^2\vec{q}_{1\perp}d^2\vec{q}_{2\perp}d^2\vec{q}_{3\perp}}{(2\pi)^9}\delta^{(2)}(\vec{q}_{1\perp}+\vec{q}_{2\perp}+\vec{q}_{3\perp})$.
In this configuration, the three quarks only carry longitudinal momenta to form the nucleon state. The above parameterization applies to both initial and final state nucleons. Of course, their momenta are different. In addition, because the quark helicities are conserved, the nucleon helicity is also conserved.

\subsection{Partonic Scattering Amplitude}

Schematically, we can write the helicity-conserved amplitude as
\begin{align}
    {\cal A}_3 &=\langle J/\psi(\epsilon_\psi),N'_\uparrow|  \gamma(\epsilon_\gamma),N_\uparrow\rangle 
    \notag\\
    &=\int [d x][d y] \Phi(x_1,x_2,x_3)\Phi^*(y_1,y_2,y_3)\nonumber\\
    &~~\times {\cal M}_\psi^{\mu\nu}(\epsilon_\gamma,\epsilon_{\psi})\frac{1}{(-t)^2} 
    {\cal H}^{\mu\nu}(\{x\},\{y\})\ , \label{eq:twist33}
\end{align}
where $\{x\}=(x_1,x_2,x_3 )$ represent the momentum fractions carried by the three quarks, $[d x]= d x_1 d x_2 d x_3\delta(1-x_1-x_2-x_3)$, and $ \Phi_3(x_1,x_2,x_3)$ is the twist-three distribution amplitude of the proton~\cite{Lepage:1980fj,Braun:1999te}. The partonic amplitude ${\cal H}^{\mu\nu}$ is calculated from the lower part of Fig.~\ref{fig:twogluon}, where the incoming three quarks carry momenta of $x_1p_1$, $x_2p_1$ and $x_3p_1$ and outgoing quarks with momenta of $y_1p_2$, $y_2p_2$ and $y_3p_2$, respectively. 

There are total of 12 diagrams (lower part) for the ${\cal H}^{\mu\nu}$. However, all the diagrams can be generated by only two specific diagrams with different helicity configurations and arrangement (permutation) of the momenta for the quark lines.  First, all these diagrams have the same color factor,
\begin{align}
C_B^2&\equiv\delta^{ac}\frac{1}{6}\epsilon_{ijk}\epsilon_{i'j'k'} (T^a)_{i'i} (T^c T^b)_{j'j}(T^b)_{k'k}
\notag \\&=\left (\frac{2}{3}\right)^2.
\end{align}
For these diagrams, a pair of quarks has zero total helicity. One can combine these two fermion lines into a Dirac trace,  by applying the following identity,
\begin{align}
&\bar U_{\uparrow/\downarrow}(p_2)  \Gamma U_{\uparrow/\downarrow}(p_1) =\bar U_{\downarrow/\uparrow}(p_1)  \Gamma_R U_{\downarrow/\uparrow}(p_2) \ , 
\end{align}
where $\Gamma_R$ is a $\gamma$-matrix chain obtained by reversing the order in $\Gamma$. This leads to the typical Dirac algebra for the partonic amplitude ${\cal H}^{\mu\nu}$,
\begin{align}
\bar U_\uparrow(p_2)  \Gamma_1 U_\uparrow(p_1) 
\ \bar U_\uparrow(p_1)  \Gamma_{2R} U_\uparrow(p_2) 
\ \bar U_\uparrow(p_2)  \Gamma_3 U_\uparrow(p_1) .
\end{align}
It is easy to find out that the first two factors can be combined into a Dirac trace, and we obtain the following expression,
\begin{align}
\text{Tr} \bigg[\frac{1+\gamma_5}{2}\slashed p_2 \Gamma_1 \frac{1+\gamma_5}{2}\slashed p_1 \Gamma_{2R} \bigg]
\ \bar U_\uparrow(p_2)  \Gamma_3 U_\uparrow(p_1) \ .
\end{align}
We will apply the above simplification to all the diagrams.

Furthermore, by examining the two gluon kinematics, we realize that one of the gluons' kinematics is determined completely by the quark line that the gluon attaches. Let us identify that gluon is ``$k_1$". Therefore, $k_1=x_ip_1-y_ip_2$ where $i$ represents the quark line in the diagram. With $k_1$ determined, we immediately deduce that $k_2=\bar x_i p_1-\bar y_i p_2$. 

Therefore, we can classify the partonic scattering amplitudes into two groups: $k_1=x_ip_1-y_ip_2$ attaches to the helicity-up quark line (Type-I) and $k_1$ attaches to the helcity-down quark line (Type-II). The derivations for both types are similar but differ in some details. 

\begin{figure}[tpb]
\includegraphics[width=0.95\columnwidth]{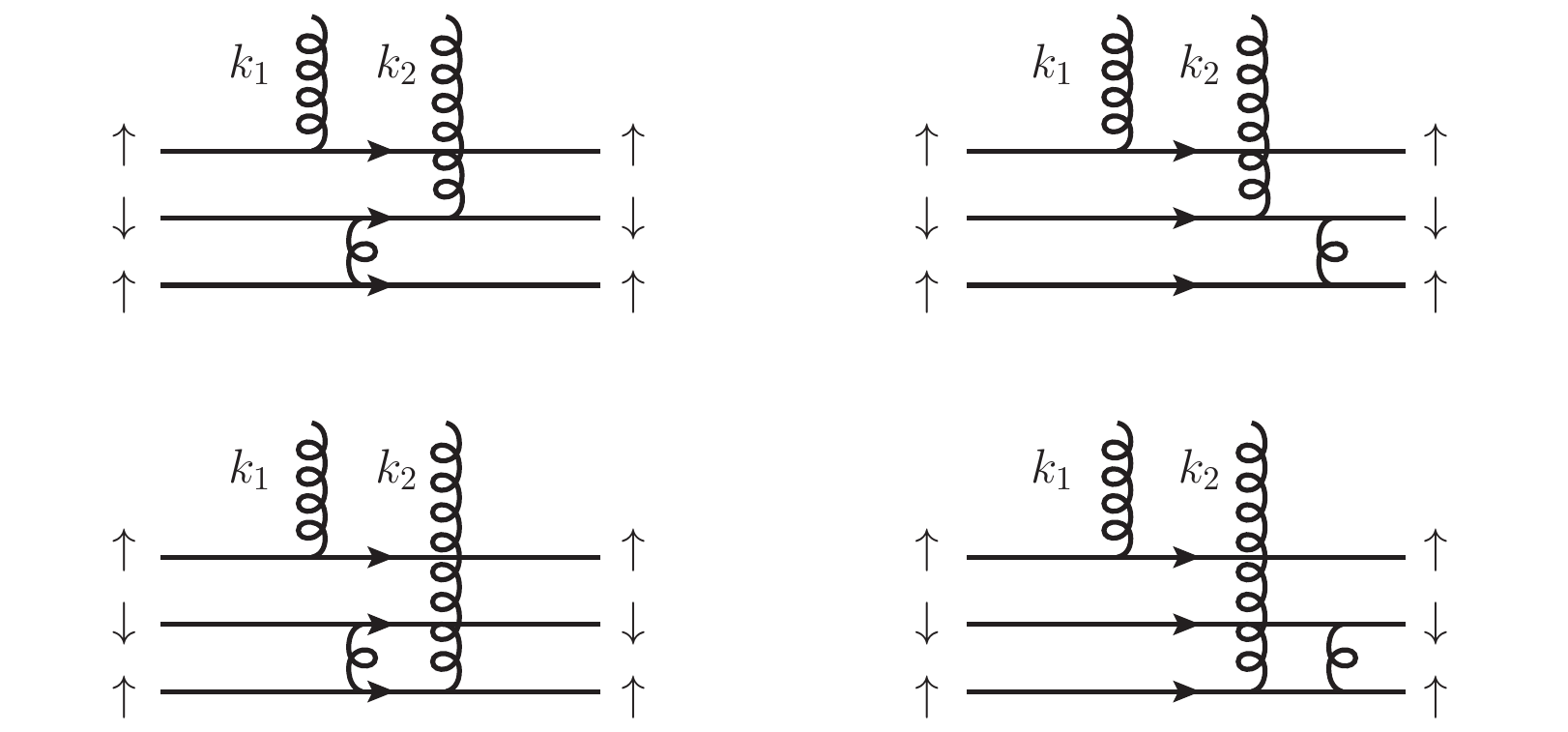}
    \caption{Partonic scattering amplitude for the type I configuration: $k_1$ is determined by the quark line with helicity-up state.}
    \label{fig:parton1}
\end{figure}

The typical diagrams of Type-I are shown in Fig.~\ref{fig:parton1}, where we include all possible attachments of $k_2$ and the additional gluon exchange between the two quark lines. The contributions of all these four diagrams can be evaluated at the same time and will be grouped together. For these diagrams, it is easy to show that the amplitude can be written as,
\begin{equation}
\bar U_\uparrow(p_2)\gamma^\mu U_\uparrow(p_1){\rm Tr}\left[\frac{1+\gamma_5}{2}\slashed{p}_2\cdots \gamma^\nu\cdots \frac{1+\gamma_5}{2}\slashed{p}_1 \cdots\right] \ .
\end{equation}
Because there is no other vector than $p_1$, $p_2$ and $\nu$, we conclude that the trace of the second factor is proportional to $p_1^\nu$ or $p_2^\nu$. Explicitly, these four diagrams contribute,
\begin{eqnarray}
&&\frac{p_2^\nu}{x_3y_3\bar x_1} \ ,~~
\frac{p_1^\nu}{x_3y_3\bar y_1} \ ,~~
\frac{p_2^\nu}{x_2y_2\bar x_1} \ ,~~
\frac{p_1^\nu}{x_2y_2\bar y_1} \ .
\end{eqnarray}
Adding them together, we have,
\begin{eqnarray}
    \frac{1}{x_1y_1\bar x_1\bar y_1}\left(\frac{1}{x_2y_2}+\frac{1}{x_3y_3}\right)\frac{\bar x_1p_1^\nu+\bar y_1p_2^\nu}{\bar x_1\bar y_1} \ ,
\end{eqnarray}
where we have also included the $t$-channel gluon propagators. 

\begin{figure}[tpb]
\includegraphics[width=0.95\columnwidth]{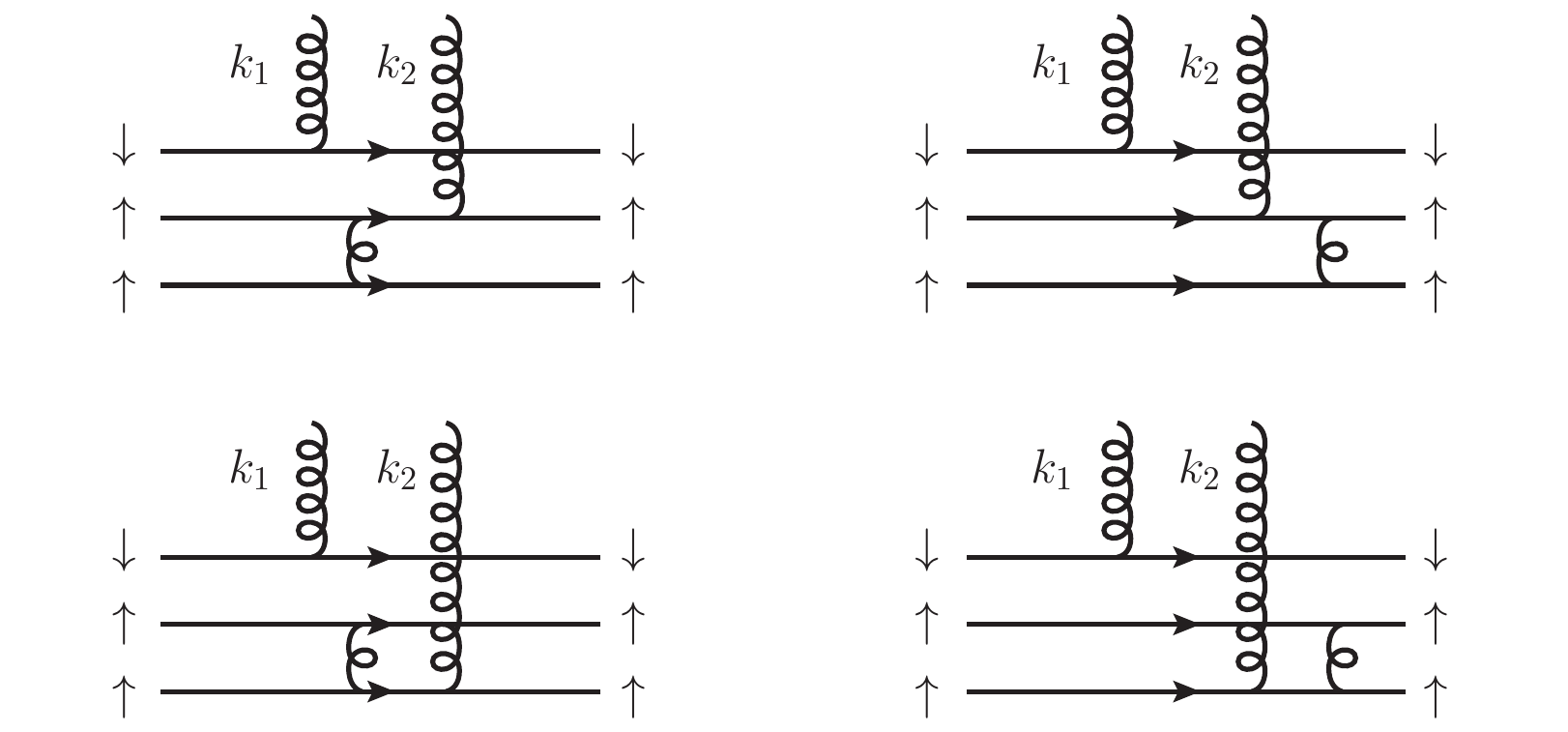}
    \caption{Partonic scattering amplitude for the type II configuration: $k_1$ is determined by the quark line with helicity-down state.}
    \label{fig:parton2}
\end{figure}

Typical Type-II diagrams are shown in Fig.~\ref{fig:parton2}. The calculations are a little bit involved. For example, the amplitude can be written in the following form,
\begin{equation}
\bar U_\uparrow(p_2)\gamma^\rho U_\uparrow(p_1){\rm Tr}\left[\frac{1+\gamma_5}{2}\slashed{p}_2\cdots \gamma^\nu\cdots \frac{1+\gamma_5}{2}\slashed{p}_1 \cdots\gamma_\rho\cdots \right] \ .
\end{equation}
Now the trace of the Gamma matrices can lead to a term like $\epsilon^{p_1p_2\nu\rho}$, which can be simplified by applying the following identity,
\begin{align}
\bar U_\uparrow(p_2) \gamma^\mu U_\uparrow(p_1)&=\frac{i\epsilon^{\mu\nu p_1p_2} }{p_1\cdot p_2} \bar U_\uparrow(p_2) \gamma_\nu U_\uparrow(p_1) \ .
\end{align}
In the end, we find that there is cancellation between different terms and the Type-II diagrams vanish.  

To derive the final result for the amplitude, we need to contract the ${\cal M}^{\mu\nu}$ with ${\cal H}^{\mu\nu}$ in Eq.~(\ref{eq:twist33}). The final results can be summarized as
\begin{align}
    {\cal A}_3 &=\langle J/\psi(\epsilon_\psi),N'_\uparrow|  \gamma(\epsilon_\gamma),N_\uparrow\rangle 
    \notag\\
    &=\int [d x][d y] \Phi(x_1,x_2,x_3)\Phi^*(y_1,y_2,y_3)\frac{1}{(-t)^2} \nonumber\\
    &~~\times \bar U_\uparrow(p_2)\slashed{k}_\gamma U_\uparrow(p_1){\cal M}^{(3)}(\epsilon_\gamma,\epsilon_{\psi},\{x\},\{y\})\ . \label{eq:twist3}
\end{align}
The spinor structure in the above equation is a consequence of the leading-twist amplitude which conserves the nucleon helicity. This is similar to the $A$ form factor calculation in Ref.~\cite{Tong:2021ctu}.

\subsection{Threshold Expansion}

In the threshold limit, we find that ${\cal M}^{(3)}$ can be further simplified as
\begin{align}
{\cal M}^{(3)}  =\epsilon^*_\psi\cdot \epsilon_\gamma\frac{8e_c e g_s^6}{27 \sqrt{3M_\psi^{7} }} \psi_{J}(0)  \left (2{\cal H}_3+{\cal H'}_3 \right)\ .\quad  
\end{align}
The coefficient ${\cal H}_3$ can be summarized as
\begin{align}
{\cal H}_3=
 I_{13}+ I_{31}+ I_{12}+I_{32}
,
\label{eq:H3}
\end{align}
where 
\begin{equation}
    I_{ij}=\frac{1}{x_i x_j y_i y_j \bar{x}_i^2 \bar{y}_i}
\end{equation}
and ${ \cal H}_3'={ \cal H}_3(y_1\leftrightarrow y_3)$.

Similar to the pion case, we can reproduce the above result by applying the GPD gluon distribution $H_g(x,\xi,t)$ at large momentum transfer in the GPD formalism. For the reference, we list the GPD gluon distribution $H_g$ in Appendix B.

The final result for the differential cross section will depend on the threshold limit of the amplitude squared. In the limit of $\chi\to 1$ we find the following result,
\begin{equation}
    |\overline{{\cal A}_3}|^2=(1-\chi)G_\psi G_{p3}(t)G_{p3}^*(t)  \ ,\label{eq:twist3p}
\end{equation}
which actually vanishes at the threshold. In the above, the spin sum and average has been performed, and $G_\psi$ has been defined in Eq.(\ref{eq:gpsi}).
$G_{p3}$ follows the form factor factorization and can be written as
\begin{equation}
    G_{p3}(t)=
    \frac{ 8\pi^2\alpha_s^2  C_B^2}{3t^2  }
    \int [dx][dy]\Phi_3(\{x\})\Phi_3^*(\{y\})\left[2{\cal H}_3+{\cal H}_3'\right] \ ,
\end{equation}
where ${\cal H}_3$ and  ${\cal H}'_3$ are given above, and $C_B^2=(2/3)^2$ is the color factor related to partonic amplitudes.
Combining $G_{p3}$ and $G_{p3}^*$, this leads to $1/(-t)^4$ power behavior for the amplitude squared, which is consistent with the conventional power counting analysis. However, this contribution is suppressed at the threshold.

The suppression factor $(1-\chi)$ comes from the spinor structure in Eq.~(\ref{eq:twist3}). In order to obtain a nonvanishing contribution at the threshold, we have to go beyond the leading-twist contributions. In the following section, we consider the three-quark Fock states with one unit OAM, which are related to the twist-four distribution amplitudes~\cite{Ji:2002xn,Braun:1999te}.

\section{Nucleon case: Twist-four contributions}

The twist-four contribution comes from the three-quark Fock state with one unit quark OAM. Two important features emerge for nonzero OAM contributions. First, the partonic scattering amplitudes conserve the quark helicities. However, because of a nonzero OAM for one of the three-quark state, the helicity of the nucleon states will be different. This contributes to the hadron helicity-flip amplitude. Second, in order to get a nonzero contribution, we have to perform the intrinsic transverse momentum expansion for the hard partonic scattering amplitudes~\cite{Belitsky:2002kj}, which will introduce an additional suppression factor of $1/(-t)$. 

The twist-four distribution amplitudes are related to the three-quark Fock states with one unit of OAM. This can comes from either the initial or final state. For example, if we consider the contribution from the initial state of spin-down nucleon, we can parameterize the Fock state as~\cite{Ji:2002xn},
\begin{eqnarray}
&&  |p_1\downarrow\rangle_{1/2} = \int d[1]d[2]d[3]\left((q_1^x-iq_1^y)
         \tilde \psi^{(3)}(1,2,3)\right.\nonumber\\
&&\left.         + (q_2^x-iq_2^y) \tilde \psi^{(4)}(1,2,3)\right) \frac{\epsilon^{ijk}}{\sqrt{6}} \nonumber \\
         &&  \times \left( u^{\dagger}_{i\downarrow}(1)
            u^{\dagger}_{j\uparrow}(2)d^{\dagger}_{k\uparrow}(3)
            -d^{\dagger}_{i\downarrow}(1)u^{\dagger}_{j\uparrow}(2)
             u^{\dagger}_{k\uparrow}(3)\right)
         |0\rangle \ , \nonumber
\end{eqnarray}
where the total quark helicity equals to $+1/2$ with nucleon helicity $-1/2$. With this choice, the final state nucleon's Fock state can be taken as that in the previous section. 

An important step in the computation of twist-four contribution is to perform the collinear expansion of the partonic scattering amplitude in terms of the transverse momenta $q_{i\perp}$. In particular, the linear term of $q_{i\perp}$ will lead to the twist-four distribution amplitudes when we integrate over the $q_{i\perp}$~\cite{Belitsky:2002kj},
\begin{eqnarray}
    &&\Psi_4(x_1,x_2,x_3) =  -\frac{2\sqrt{6}}{x_2M}\int [dq_\perp]
    \nonumber \\ 
    && ~~\times   \vec{q}_{2\perp}\cdot \left[   \vec{q}_{1\perp} \tilde \psi^{(3)}(1,2,3)
 + \vec{q}_{2\perp} \tilde \psi^{(4)}(1,2,3)\right] \ ,\label{eq:q2}\\
&&    \Phi_4(x_2,x_1,x_3) =  -\frac{2\sqrt{6}}{x_3M}\int [dq_\perp] 
    \nonumber \\ 
    &&~~\times \vec{q}_{3\perp}\cdot \left[   \vec{q}_{1\perp} \tilde \psi^{(3)}(1,2,3)
 +  \vec{q}_{2\perp} \tilde \psi^{(4)}(1,2,3)\right] \ .\label{eq:q3} 
\end{eqnarray}
To extract the linear dependence of the transverse momentum $q_{i\perp}$ from the partonic amplitudes, one can first expand the spinor as
\begin{align}
U(x_i p_1+ \vec q_{i\perp})&\approx U(x_i p_1)+\frac{\slashed {\vec q}_{i\perp} \slashed p_2}{2x_i p_2  \cdot p_1 }  U(x_i p_1)\ .
\end{align}
After the evaluation of the Dirac structures in the amplitudes following the strategy in last section,  all the linear dependence of $\vec q_i$ is explicit and straightforward to find out.
For the contributions associated with the initial OAM, it will yield a structure like:
\begin{align}
&{ \Gamma}_1(\{x\},\{y\})( q_1^x+i q_1^y)
\bar U_\uparrow(p_2)  U_\downarrow(p_1)
\notag \\
&+{ \Gamma}_3(\{x\},\{y\})( q_3^x+i q_3^y) \bar U_\uparrow(p_2)  U_\downarrow(p_1)\ .
\end{align}
where the transverse momentum conservation $ \vec{q}_{2\perp }=-  \vec{q}_{1\perp}-  \vec{q}_{3\perp}$ is used, and the identities
$
\gamma^x U_\uparrow(p)=U_\downarrow(p),\  \gamma^y U_\uparrow(p)=i U_\downarrow(p)
$ have been applied.  

Applying Eqs.~(\ref{eq:q2},\ref{eq:q3}) with the linear terms of $q_{i\perp}$ from the partonic amplitudes, we obtain the twist-four contribution to the scattering process of $\gamma p\to J/\psi p$ as
\begin{eqnarray}
      {\cal A}_4&=&\langle J/\psi(\epsilon_\psi),N'_\uparrow|  \gamma(\epsilon_\gamma),N_\downarrow\rangle 
    \notag\\
&=&\bar U_\uparrow(p_2) U_\downarrow(p_1) \frac{M_p}{(-t)^3}\int [dx][dy]\Phi_3^*(\{y\})\nonumber\\
&&\times \left[\Psi_4(\{x\}){\cal M}_\Psi^{(4)}
+\Phi_4(\{x\}){\cal M}_\Phi^{(4)}
\right]\ ,\label{eq:twist4}
\end{eqnarray}
where $\Psi_4$ and $\Phi_4$ are the twist-four distributions introduced above and ${\cal M}_{\Psi,\Phi}^{(4)}$ from the partonic amplitudes.  From this equation, we can clearly see that the nucleon helicity-flip is manifest in the spinor structure. This amplitude is negligible at high energy, but will be important at the threshold, because it is not suppressed in the limit of $\chi\to 1$. The amplitude squared along with the associated spin sum and average can be written as
\begin{eqnarray}
    |\overline{{\cal A}_4}|^2=\widetilde m_t^2 G_\psi G_{p4}(t) G_{p4}^*(t) \ , \label{eq:twist4p}
\end{eqnarray}
where $\widetilde m_t^2=M_p^2/(-t)$, $G_\psi$ is the same as above. $G_{p4}$ depends on the twist-three and twist-four distribution amplitudes \cite{Braun:1999te,Braun:2000kw},
\begin{align}
G_{p4}(t) &= \frac{C_B^2(4\pi\alpha_s)^2}{12t^2}  \int [d x][d y] \Phi_3(y_1,y_2,y_3)\nonumber\\
&\times \left\{ x_3 \Phi_4(x_1,x_2,x_3) { T}_{4\Phi } (\{x\},\{y\})\right.\nonumber\\
&\left.+x_1\Psi_4(x_2,x_1,x_3) { T}_{4\Psi } (\{x\},\{y\})
\right\}\   \ ,\label{eq:gp4}
\end{align} 
where the hard function has the following form
\begin{align}
&{ T}_{4\Psi} =2{\cal T}_{4\Psi}+ {\cal T}'_{4\Psi}\ ,
\notag \\
&  { T}_{4\Phi} =2{\cal T}_{4\Phi}+ {\cal T}'_{4\Phi}\ ,
\end{align}
and ${\cal T}'_4$ is obtained from ${\cal T}_4$ by interchanging $y_1$ and $y_3$. Then we have
\begin{align}
{\cal T}_{4\Psi}=&x_3 K_1 
( 1+y_2/\bar y_1)+
2\bar x_3 \tilde K_1 
\notag \\&
+
2x_3 (\tilde K_2- K_2) 
-K_3/\bar y_1 
\notag \\&
+x_3 (K_4 + K_5)/\bar x_1
+
2(\tilde K_4+\tilde K_5)~,
\notag \\
{\cal T}_{4\Phi}=&{\cal T}_{4\Psi}(1\leftrightarrow 3)~,
\end{align}
where the functions $K_i$ and $\tilde K_i$ are defined as 
\begin{align}
&K_1=\frac{1}{x_1 x_3^2 y_1 y_3^2 \bar{x}_1^2 \bar{y}_1}~,
\quad
K_2=\frac{1}{x_1 x_2 x_3^2 y_2 y_3^2 \bar{x}_2 \bar{y}_2}~,\nonumber\\
&K_3=\frac{1}{x_1 x_2 y_1 y_2 \bar{x}_1^2 \bar{y}_1}
~,
\quad
K_4=\frac{1}{x_1 x_3^2 y_1 y_3 \bar{x}_1 \bar{y}_1^2}~,
\notag \\
&K_5=\frac{1}
{x_1 x_2 x_3 y_1 y_2 \bar{x}_1 \bar{y}_1^2}~, \quad  \tilde K_i=K_i(1\leftrightarrow3)~.
\label{eq:K}
\end{align}
As mentioned above, the twist-four distribution amplitudes can come from both initial and final state nucleons. Because of the symmetric property of the partonic scattering amplitudes, these two contributions are the same and have been included in the above final result.

Eqs.~(\ref{eq:twist4p}) and (\ref{eq:twist3p}) are the final results of our analysis. Comparing these two, we find that the twist-four contribution is suppressed in $1/t$ but enhanced at the threshold. These two features can be used to disentangle their contributions in experiments. If we limit our discussions in the threshold region, the only contribution comes from the twist-four term.

\section{Interpretation in Terms of Gravitational Form Factor?}

As mentioned in Introduction, the near threshold heavy quarkonium production has been argued to provide a direct access to the gluonic gravitational form factors of the nucleon. However, our explicit calculations for the pion case have shown that there is no direct connection between them. 

From the results in previous sections, we have calculated the near threshold photo-production of heavy quarkonium on the nucleon target at large momentum transfer. The gluonic form factors at large $(-t)$ have been recently calculated in Ref.~\cite{Tong:2021ctu}. We conclude, again, we can not build a direct connection between them. 

\subsection{Construct the Gluonic Operator}

The above conclusion can be understood from a detailed analysis of the photon-quarkonium transition amplitude. As discussed in Sec.~II, this amplitude can be simplified in the heavy quark mass limit, $M_V^2\gg (-t)$, 
\begin{eqnarray}
    {\cal M}_\psi^{\mu\nu}=N_\psi \epsilon_\psi^*\cdot \epsilon_\gamma\frac{k_{\gamma,\alpha }k_{\gamma,\beta}}{k_1\cdot k_\gamma k_2\cdot k_\gamma}{\cal W}_T^{\alpha\beta\mu\nu} \ .
\end{eqnarray}
Here, we only keep the leading term in this limit. For simplicity, we have also dropped the associated color factors associated with the $t$-channel gluons. In the above, ${\cal W}_T^{\alpha\beta\mu\nu}$ is defined as
\begin{eqnarray}
    {\cal W}_T^{\alpha\beta\mu\nu}&=&-k_1^\alpha k_2^\beta g^{\mu\nu}-k_1\cdot k_2 g^{\alpha\mu}g^{\beta\nu}\nonumber\\
    &&+
    k_1^\nu k_2^\beta g^{\alpha\mu}+k_2^\mu k_1^\alpha g^{\beta \nu} \ ,
\end{eqnarray}
which can be identified as gluonic operator of ${F^{\alpha}}_{\rho}F^{\beta\rho}$ acting on the nucleon state. However, the complete scattering amplitude involves the integral of the momenta $k_1$ and $k_2$ with the associated propagators depending on them. In the end, the $\gamma N\to J/\psi N'$ amplitude can be schematically written as 
\begin{eqnarray}
{\cal A}&=&N_\psi \epsilon_{\psi}^*\cdot \epsilon_\gamma \nonumber\\
&&\times \int d^4k_1d^4k_2   \frac{k_{\gamma,\alpha }k_{\gamma,\beta}}{(k_1\cdot k_\gamma-i \varepsilon) (k_2\cdot k_\gamma-i \varepsilon)}\nonumber\\
&& \times \int d^4\eta_1d^4\eta_2 e^{ik_1\cdot \eta_1+ik_2\cdot \eta_2}   \nonumber\\
&&~~~~\times \langle N'|
    {F^{a,\alpha}}_{\rho}(\eta_1)F^{a,\beta\rho}(\eta_2)|N\rangle  \ .
\end{eqnarray}
Clearly, if we neglect the $k_1$ and $k_{2}$ dependence in the pre-factor of $\frac{1}{(k_1\cdot k_\gamma-i \varepsilon) (k_2\cdot k_\gamma-i \varepsilon)} $, the above equation can be identified as a gluonic gravitational form factor of the nucleon state. However, as discussed in Sec.~II, this pre-factor comes from the quark propagators in the photon-quarkonium transition amplitude. The complete calculation will have a full dependence on the momentum fractions of the incoming nucleon $p_1$ carried by the two gluons $k_1$ and $k_2$.  

We emphasize that the above discussions apply to all of the kinematics in heavy quarkonium photo-production, including small and large $(-t)$. Therefore, our conclusion is valid in the whole kinematics of this process that there is no direct connection between the near threshold photo-prodution of heavy quarkonium and the gluonic gravitational form factors of the nucleon.

\subsection{Compare to the GPD Formalism}

It is interesting to find out that the above ${\cal W}_T^{\mu\nu}$ can be directly compared to that for the gluon GPD calculations. Gluon GPD is defined through the matrix element $\langle N'|F^{+\alpha}F^{+}_{\ \alpha}| N\rangle$. The amplitude associated with this can be written as
\begin{equation}
    -n\cdot k_1 n\cdot k_2g^{\mu\nu}-n^\mu n^\nu k_1\cdot k_2
    +n^\mu k_1^\nu n\cdot k_2+n^\nu k_2^\mu n\cdot k_1 \ ,
\end{equation}
where $n$ is the light-cone vector used in the GPD definition with $n\cdot k=k^+$ for any momentum $k$. Here, $k_1$ and $k_2$ represent the gluon momenta that couple to the nucleon state, and $\mu\nu$ for their polarization indices. Clearly, this is the same structure as ${\cal W}_T^{\mu\nu}$ of previous subsection if we identify $n\propto k_\gamma$.

Following this argument, the scattering amplitude of $\gamma N\to J/\psi N'$ can be formulated in terms of the gluon GPDs~\cite{Hoodbhoy:1996zg,Ivanov:2004vd,Koempel:2011rc,Boussarie:2020vmu,Hatta:2021can,Guo:2021ibg},
\begin{eqnarray}
  &&  {\cal A}=N_\psi \epsilon_\psi^*  \cdot  \epsilon_\gamma \int_{-1}^{1} dx \frac{1}{(x+\xi-i\varepsilon)(x-\xi+i\varepsilon)}\\
    &\times & \frac{1}{\bar P^+}\int\frac{d\eta^-}{2\pi}e^{ix\bar P^+\eta^-}\langle N'|
    {F^{a,+}}_{\alpha}(-\frac{\eta^-}{2})F^{a,\alpha+}(\frac{\eta^-}{2})|N\rangle \ ,\nonumber
\end{eqnarray}
where the last factor defines the associated gluon GPDs. In previous sections, we have given explicit examples that demonstrate the consistency between our calculations with the GPD formalism.

Clearly, from the above GPD formalism, one can only link to the gluonic gravitational form factors by making approximations of no $x$-dependence in the pre-factor $\frac{1}{(x+\xi-i\varepsilon)(x-\xi+i\varepsilon)}$~\cite{Hatta:2021can,Guo:2021ibg}. This is the same as we discussed in the previous subsection. Therefore, our conclusion of no direct connection between the near threshold photo-production of heavy quarkonium state and the gluonic gravitational form factors is consistent with the GPD formalism.

\section{Phenomenology Applications}

Taking into account the contributions derived in previous sections, we can write down the differential cross section for the near threshold heavy quarkonium photoproduction at large momentum transfer,
\begin{eqnarray}
\frac{d\sigma}{dt}|_{(-t)\gg\Lambda_{QCD}^2}&=&\frac{1}{16\pi(W_{\gamma p}^2-M_p^2)^2}\left( |\overline{{\cal A}_3}|^2+|\overline{{\cal A}_4}|^2\right)\nonumber\\
&\approx &\frac{1}{(-t)^4}\left[(1-\chi){\cal N}_3+\widetilde m_t^2 {\cal N}_4\right]\ ,\label{eq:diff}
\end{eqnarray}
where ${\cal N}_3$ and ${\cal N}_4$ represent the twist-three and twist-four contributions, respectively. The most important consequence of our power counting analysis is that the leading-twist contribution is suppressed at the threshold. Away from the threshold point, it will start to contribute and may dominate at large $(-t)$ because of the leading power feature. With high precision future experiments~\cite{Chen:2014psa,AbdulKhalek:2021gbh,Joosten:2018gyo}, we should be able to distinguish their contributions. 

If we take the leading contribution of Eq.~(\ref{eq:diff}) at the threshold, i.e., the ${\cal N}_4$ term, the differential cross section only depends on the momentum transfer $t$. This is an important signal from the perturbative QCD analysis in this paper. Of course, away from the threshold region, we have to take into account additional contribution from ${\cal N}_3$ and the kinematic corrections in Eq.~(\ref{eq:diff}). In Ref.~\cite{Sun:2021gmi}, the twist-four contribution has been applied to fit the GlueX data~\cite{Ali:2019lzf} with the following parameterization of the differential cross section,
\begin{eqnarray}
\frac{d\sigma}{dt} |^{twist-4}=\frac{N_0}{(-t+\Lambda^2)^5}\ ,\label{eq:jpsi}
\end{eqnarray}
where $\Lambda^2=1.41\pm 0.20~{\rm GeV}^2$ and $N_0=51\pm 22 ~{\rm nb * GeV^8}$. The current data from the GlueX can be well described by the above parameterization. In the following, we will apply this result to the future experiments for the threshold photo-production of $\psi'$ and $\Upsilon$. 

We have also made an order of magnitude estimate of the differential cross section by applying the twist-four contribution of Eq.~(\ref{eq:twist4p}) with model assumptions for the twist-three and twist-four distribution amplitudes of the nucleon~\cite{Braun:1999te, Braun:2000kw}. There have been great efforts to compute these distribution amplitudes from various methods~\cite{Ioffe:1981kw,Chung:1981cc,Chernyak:1983ej,Chernyak:1984bm,King:1986wi,Chernyak:1987nt,Chernyak:1987nu,Bolz:1996sw,Braun:2001tj,Braun:2006hz,Gockeler:2008xv,Braun:2008ia, QCDSF:2008qtn,Passek-Kumericki:2008uqr,Lenz:2009ar,Anikin:2013aka, Braun:2014wpa,Bali:2015ykx, RQCD:2019hps}, including the lattice QCD, the light-cone sum rule and model calculations. The differential cross sections calculated from the distribution amplitudes with realistic model assumptions, e.g., those from Ref.~\cite{Braun:2006hz}, are consistent with the experimental data around $-t=1.5\rm GeV^2$ and the fitted result of Eq.~(\ref{eq:jpsi})~\footnote{In the numeric calculation of the $G_{p4}(t)$ in Eq.~(\ref{eq:gp4}), a lower cutoff ($\sim (0.17{\rm GeV})^2/(-t)$) on the momentum fractions $x_i$ and $y_i$ in the integral is imposed to avoid the end-point singularity. This is similar to the Pauli form factor calculation at large momentum transfer in Ref.~\cite{Belitsky:2002kj}.}, whereas the results from the asymptotic distribution amplitudes are an order of magnitude smaller. 

\subsection{Predictions for $\psi'$ and $\Upsilon$(nS) Production}

\begin{figure}[tpb]
\includegraphics[width=0.49\columnwidth]{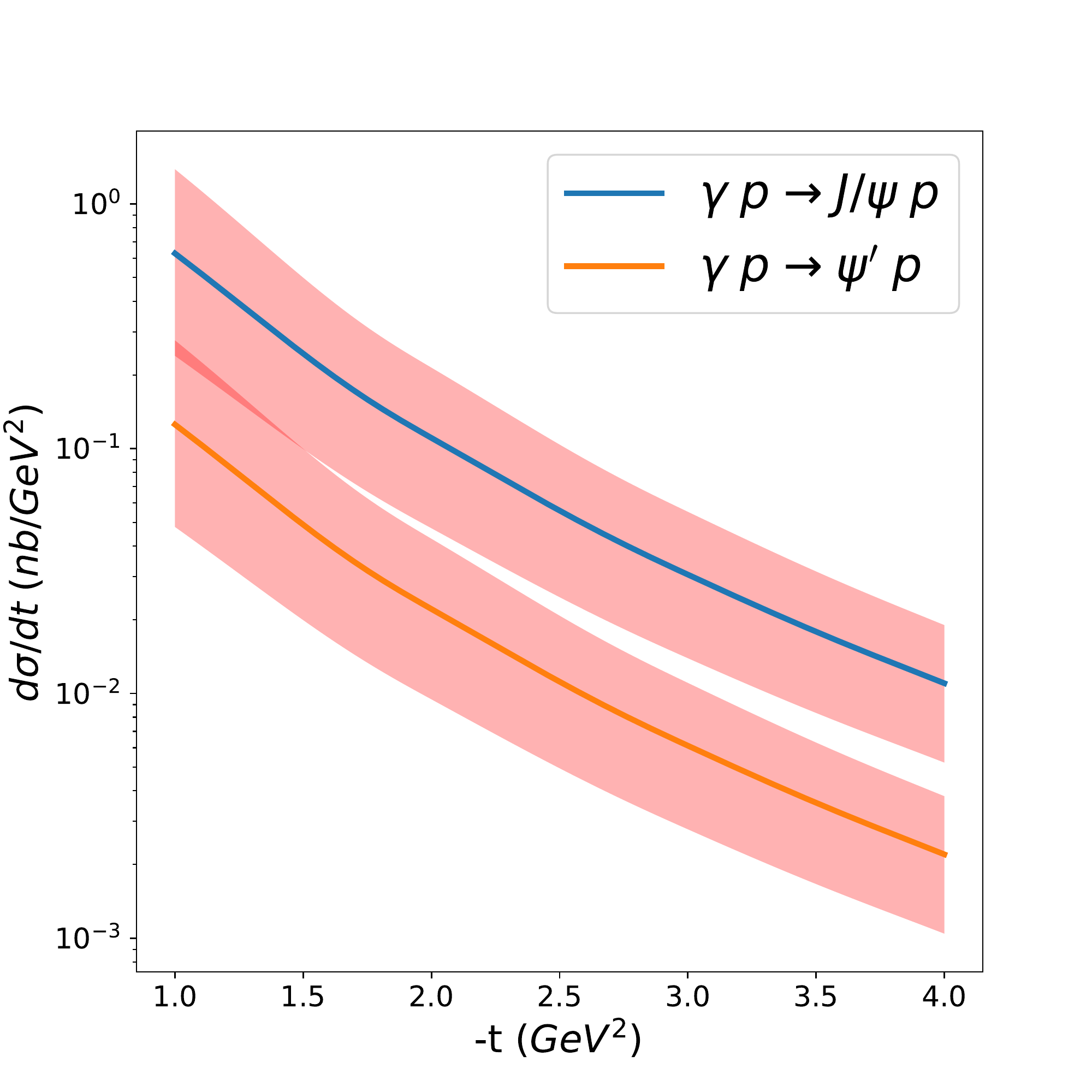}
\includegraphics[width=0.49\columnwidth]{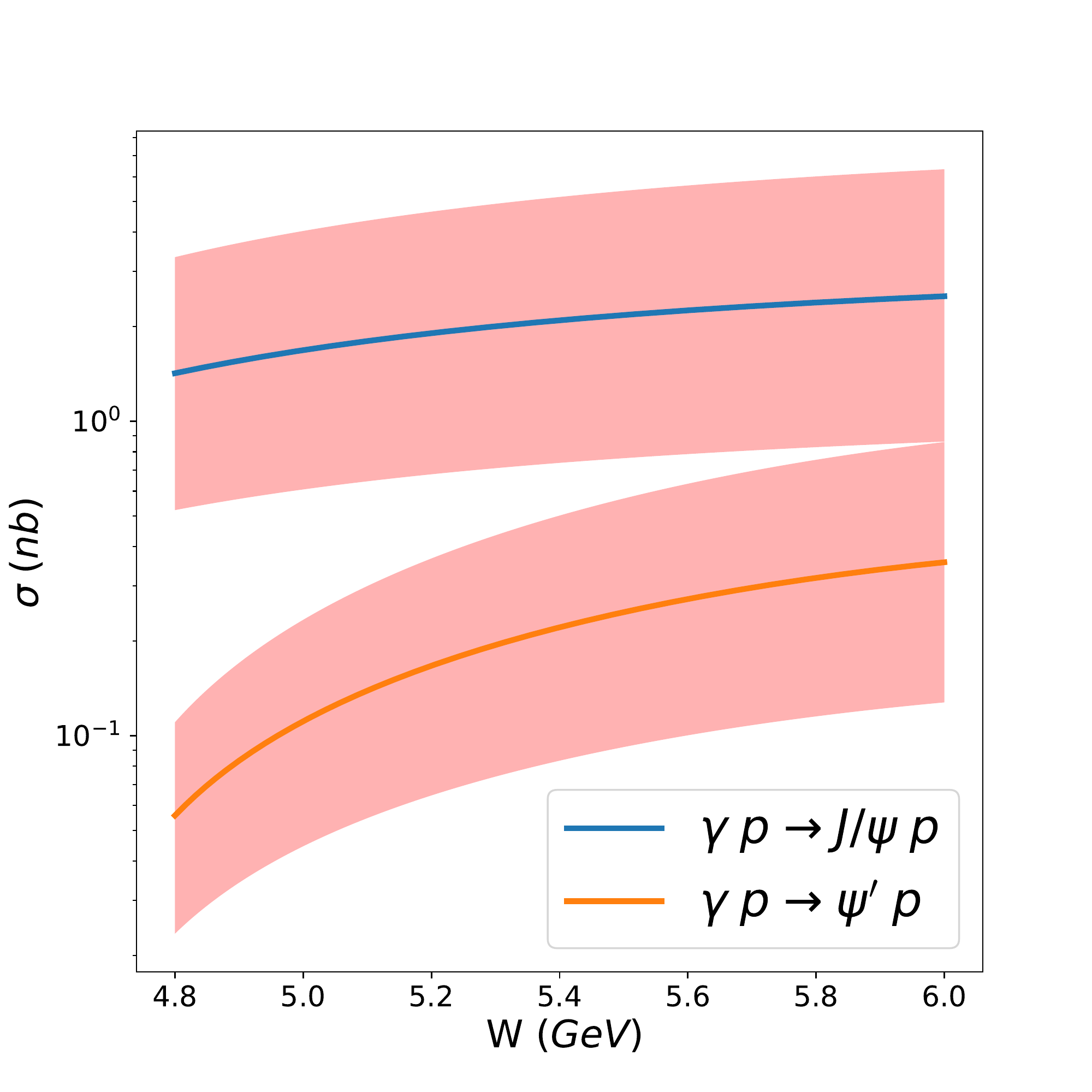}
    \caption{Differential cross sections for $J/\psi$ and $\psi'$ photo-production as functions of the momentum transfer $t$ and the total cross sections near the threshold as functions of $W_{\gamma p}$.}
    \label{fig:psi}
\end{figure}

\begin{figure}[tpb]
\includegraphics[width=0.49\columnwidth]{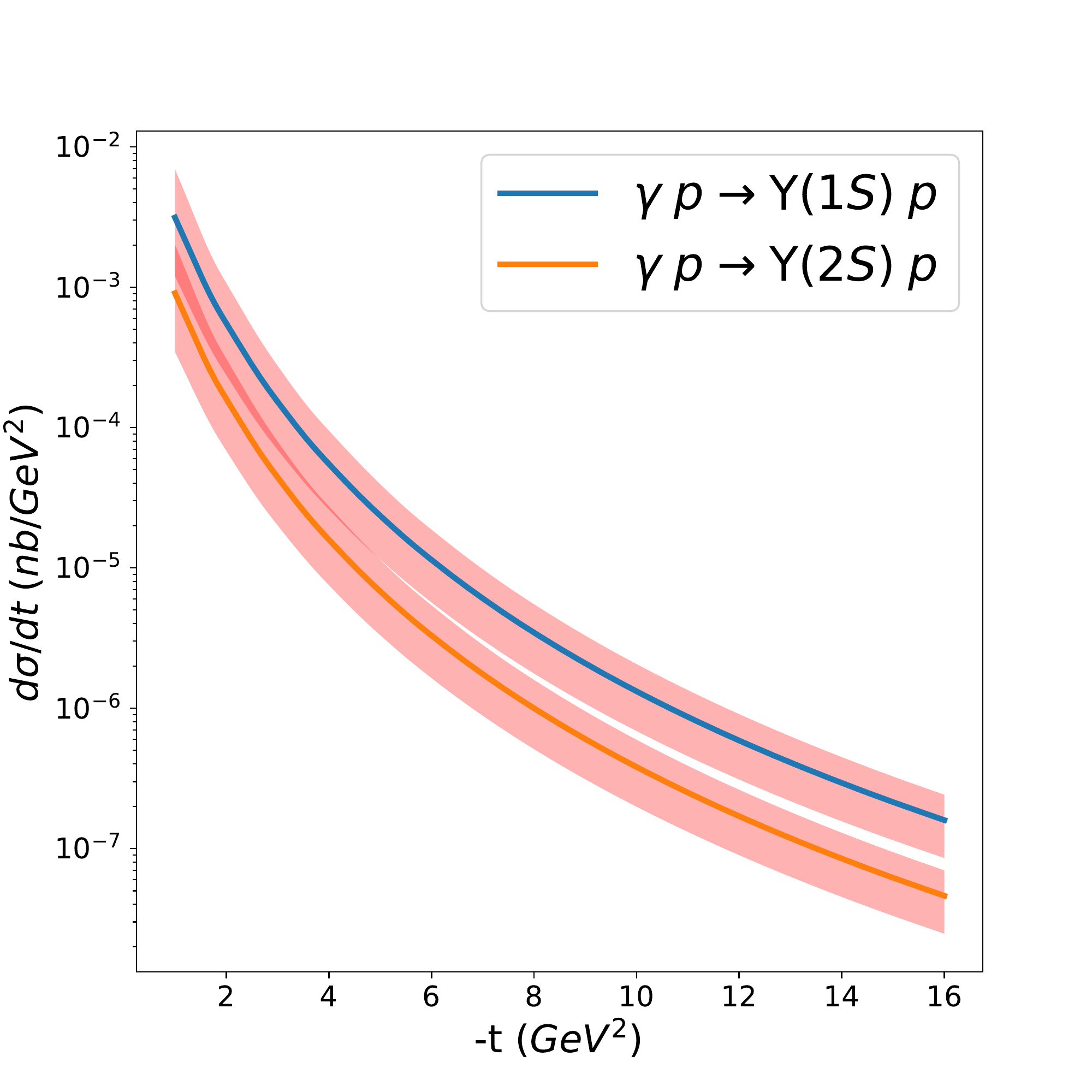}
\includegraphics[width=0.49\columnwidth]{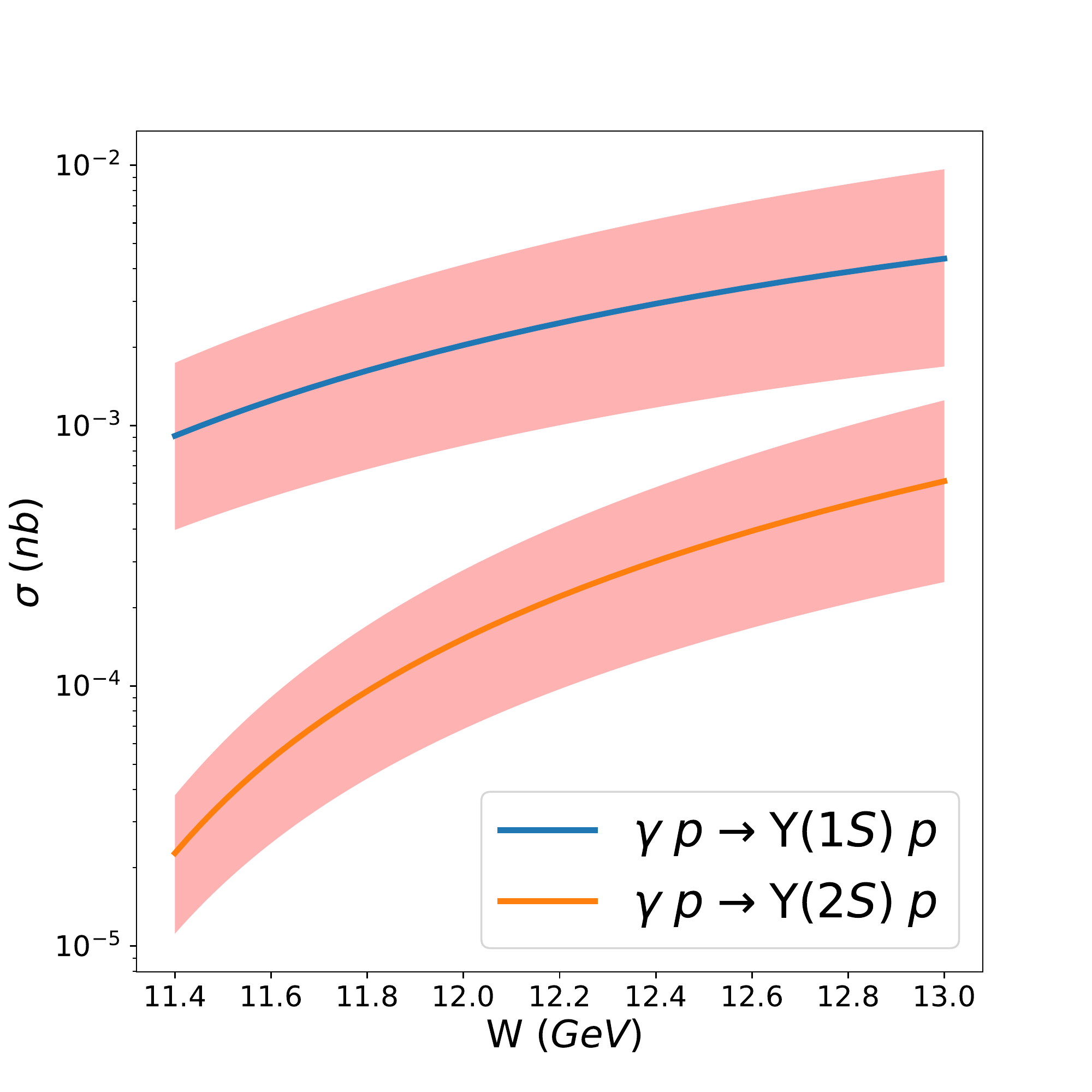}
    \caption{The differential cross sections for $\Upsilon$(1S) and $\Upsilon$(2S) photo-production as functions of the momentum transfer $t$ and the total cross sections as functions $W_{\gamma p}$ near the threshold.}
    \label{fig:upsilon}
\end{figure}

Extending our analysis to other heavy quarkonium states is straightforward and similar formulas can be derived. As a first step, we take the differential cross section from the twist-four contribution at the threshold,
\begin{eqnarray}
\frac{d\sigma(\gamma p\to Vp)}{dt}|_{threshold}=\frac{N_0^V}{(-t+\Lambda^2)^5}\ ,
\end{eqnarray}
for a heavy quarkonium state $V$. In the heavy quark mass limit, the $t$-dependence only comes from the nucleon side. Therefore, we will assume the above $\Lambda$ parameter should be same for all heavy quarkonium states. On the other hand, the normalization factor $N_0^V$ will depend on the quarkonium state in the final state. From the derivations in previous section, we know that the differential cross section is proportional to,
\begin{equation}
    \frac{d\sigma}{dt}\propto \frac{\alpha_s^2(M_V)\langle  0 | {\cal O}^{V}({}^3S_1^{(1)}) |0\rangle}{M_V^7} \ ,
\end{equation}
from which we derive the ratio between different heavy quarkonium states,
\begin{equation}
    \frac{N_0^V}{N_0}=\frac{\alpha_s^2(M_V)\langle  0 | {\cal O}^{V}({}^3S_1^{(1)}) |0\rangle/M_V^7}{\alpha_s^2(M_\psi)\langle  0 | {\cal O}^{\psi}({}^3S_1^{(1)}) |0\rangle/M_\psi^7} \ .
\end{equation}
Substituting the associated NRQCD matrix elements for $J/\psi$, $\psi'$, and $\Upsilon$ (1S, 2S) from, e.g., Refs.~\cite{Feng:2015sjx,Zhan:2021dlu}, we find the following values for the normalization factors,
\begin{eqnarray}
 &&   {N_0^{\psi'}}=0.20{N_0} \ , \\
 &&   {N_0^{\Upsilon(1S)}}=5\times 10^{-3} {N_0}\ , \\
 &&   {N_0^{\Upsilon(2S)}}=2.5\times 10^{-3} {N_0}\ .
\end{eqnarray}
In Fig.~\ref{fig:psi}, we show the threshold cross sections for $\gamma p\to \psi' p$. As comparison, we also show the results for $J/\psi$. For Upsilon production, the results are plotted in Fig.~\ref{fig:upsilon}.

The comparison between different quarkonium states will provide an important confirmation for the production mechanism. The comparison between Charmonium and Bottomonium, in particular, will test the heavy quark limit we have employed in this paper.  Meanwhile, the momentum transfer range is much higher for $\Upsilon$ as compared to $J/\psi$. This provides a unique opportunity to explore the large momentum transfer region.

\section{Conclusion}
In this paper, we have carried out a detailed derivation of near threshold heavy quarkonium photoproduction at large momentum transfer. We have taken into account the three-quark Fock state of the nucleon with zero and one unit of quark OAM. We found that the contribution from the Fock state with zero quark OAM is suppressed at threshold. The differential cross section is dominated by the contribution from nonzero OAM Fock state and has a power behavior of $1/(-t)^5$.

Our power counting predictions are consistent with recent experimental data of near threshold photo-production of $J/\psi$ from the GlueX collaboration at JLab. Based on the comparison between our derivation and the experimental data, we have made predictions for $\psi'$ and $\Upsilon$(1S,2S). All these predictions can be tested at future facilities including the electron-ion colliders~\cite{Accardi:2012qut,AbdulKhalek:2021gbh,Anderle:2021wcy}.

We have also shown that there is no direct connection between the near threshold photo-production of heavy quarkonium state and the gluonic gravitational form factors. The indirect connection can be built through GPD gluon distributions. For example, we can parameterize the gluon GPDs and fit to the experimental data, which, in return, can constrain the associated gravitational form factors.

{\bf Acknowledgments:} We thank Yoshitaka Hatta, Xiangdong Ji, and Nu Xu for discussions and comments. This material is based upon work supported by the LDRD program of Lawrence Berkeley National Laboratory, the U.S. Department of Energy, Office of Science, Office of Nuclear Physics, under contract numbers DE-AC02-05CH11231. P. Sun is supported by Natural Science Foundation of China under grant No. 11975127 and No. 12061131006 as well as Jiangsu Specially Appointed Professor Program. X. B. Tong is supported by the CUHK-Shenzhen university development fund under grant No. UDF01001859.

\newpage
\appendix
\section{Gluon GPD for Pion}
The gluon GPD for pion is defined as
\begin{align}
&\int \frac{d\eta^- }{2 \pi }   e^{ix \bar P^+ \eta^-}
\Big\langle p_2\Big|
 F^{+\mu}_a(-\frac{\eta^-}{2} ) 
{\cal L} _{ab}[-\frac{\eta^-}{2}, \frac{\eta^-}{2}]
  \notag \\ 
 & \quad \quad  \times F^{\ +}_{\mu,b}(\frac{\eta^-}{2})  
  \Big| p_1\Big\rangle
  \notag \\ 
  =& \bar P^+ H^{(\pi)}_g(x,\xi, t)~,
\end{align}
where the gluon field strength tensor is $F^a_{\mu\nu}=\partial_\mu A^a_\nu-\partial_\nu A^a_\mu-g_s f^{abc} A^b_\mu A^c_\nu$,
and the gauge link in the adjoint representation is 
\begin{align}
    {\cal L}_{ab} \left[z_2,z_1\right] ={\cal P}
  \ \text{exp} \left[ g_s  \int ^{z_2}_{z_1} \  d z^- \  G^{+,c}(z^- ) f^{acb}  \right]~.
 \end{align} 
${\cal P}$ denote the path-ordering operation. In the definition of GPD, $\bar P=(p_1+p_2)/2$ is the average momentum, $\Delta =p_2-p_1$ is the momentum transfer and $t=\Delta^2$. The skewness parameter $\xi$ is defined as the projection of the momentum transfer $ \Delta$ along $\bar P $ direction, $
\xi=-\frac{ \Delta^+ }{2  \bar P^+}.
$
\par 
In the large $(-t)$ limit, the gluon GPD of the pion has the following factorization formula:  
\begin{align}
H^{(\pi)}_g(x,\xi, t)
=&\int d x_1 d y_1 \phi(y_1)\phi(x_1) {\cal H }^{(\pi)}(x_1,y_1)~ \ ,
\end{align}
where $\phi$ represents the leading-twist distribution amplitude of pion. The perturbative function at the leading order can be written as
\begin{align}
{\cal H }^{(\pi)}(x_1,y_1)=&\frac{g_s^2 C_F }{-t} 
 \left(\frac{(1-\xi)^2}{x_1 \bar x_1}+\frac{(1+\xi)^2}{y_1 \bar y_1} 
\right )
\notag \\ 
& \times \bigg( \delta\big[ x- (x_1-y_1+ \xi(x_1+y_1-1))\big]\notag\\
&+ \delta\big[ x+ (x_1-y_1+ \xi(x_1+y_1-1))\big] \bigg)~.
\end{align}
The above result is similar to that of the quark GPD calculated in Ref.~\cite{Hoodbhoy:2003uu} for the pion.


\section{Gluon GPD for Nucleon}
The  gluon GPD of nucleon is defined from 
 \begin{align}
&\int \frac{d\eta^- }{2 \pi }   e^{i x \bar P^+ \eta^- } 
\Big\langle p_2 ,s'\Big|
 F^{+\mu}_a(-\frac{\eta^-}{2}) 
{\cal L}_{ab}[-\frac{\eta^-}{2}, \frac{\eta^-}{2}]
\notag \\ 
 & \quad  \times  F^{\  +}_{\mu,b}(\frac{\eta^-}{2})  
  \Big| p_1,s \Big\rangle
  \notag\\
 =&  \frac{ 1}{2} \bigg( H_g(x,\xi, t) \bar U(p_2,s') \gamma^+ U(p_1,s)
 \notag \\ 
 &\quad+E_g(x,\xi,t)\bar U(p_2,s') \frac{i \sigma^{+ \alpha} \Delta_\alpha}{2M_p}U(p_1,s) \bigg)~,
 \end{align}
  where  $s^\mu$ denote the covariant spin-vector of the proton. 
\par 
Following the strategy  in \cite{Hoodbhoy:2003uu,Tong:2021ctu}, the GPD $H_g$ can be extracted from the helicity conserved amplitude, and one can show that at the large momentum transfer, $H_g$ follows the following factorization formula:
\begin{align}
H_g(x,\xi, t) =&\int [d x][ dy]\Phi_3^*(y_1,y_2,y_3)
  \Phi_3(x_1,x_2,x_3)
 \notag \\ &\quad\quad  \times {\cal H }(\{x\} ,\{y\})~,
 \end{align}
where $\Phi_3$ is the twist-3 proton light-cone amplitude~\cite{Braun:1999te}, and $ {\cal H }$ is the hard coefficient and perturbatively calculable.  At the leading order, we obtain
 \begin{align}
  {\cal H }(\{x\} ,\{y\}) &=2\tilde   {\cal H }+\tilde   {\cal H }(y_1 \leftrightarrow y_3)~,
\end{align} 
where
\begin{eqnarray}
\tilde   {\cal H }
&&=\frac{ 4\pi^2\alpha_s^2  C_B^2}{3t^2  }\times
\notag \\
&&\bigg\{
  \bigg(\frac{x_1+y_1+ \xi(x_1-y_1)}{\bar x_1 \bar y_1 x_1  x_3 y_1  y_3 } +\frac{x_1+y_1+ \xi(x_1-y_1)}{\bar x_1 \bar y_1 x_1  x_2 y_1  y_2 } \bigg)\notag\\
&&\times
\bigg( \delta\big[ x- (x_1-y_1+ \xi(x_1+y_1-1))\big]\notag\\
&&+ \delta\big[ x+ (x_1-y_1+ \xi(x_1+y_1-1))\big] \bigg) 
\notag\\ 
&& +\bigg( \frac{x_3+y_3+ \xi(x_3-y_3)}{\bar x_3 \bar y_3 x_3  x_1 y_3  y_1 } +\frac{x_3+y_3+ \xi(x_3-y_3)}{\bar x_3 \bar y_3 x_3  x_2 y_3  y_2 } \bigg) \notag\\
&&\times\bigg( \delta\big[ x- (x_3-y_3+ \xi(x_3+y_3-1))\big]\notag\\
&&+ \delta\big[ x+ (x_3-y_3+ \xi(x_3+y_3-1))\big] \bigg) 
\bigg\}~.
\end{eqnarray}
Similar results for the quark GPDs $H_q$ of the nucleon have been calculated in Ref.~\cite{Hoodbhoy:2003uu}. They share the same power behavior at large $(-t)$.

On the other hand, the GPD $E_g$ at large $(-t)$ is calculated from the nucleon helicity-flip amplitude, and the related factorization formula can be written as
\begin{align}
&E_g(x,\xi,t) =  \int [d x][d y] \left\{ x_3 \Phi_4(x_1,x_2,x_3) {\cal E}_{\Phi g}(\{x\},\{y\})\right.\nonumber\\
&
\left.+x_1\Psi_4(x_2,x_1,x_3) {\cal E}_{\Psi g}(\{x\},\{y\})
\right\}\  \Phi_3(y_1,y_2,y_3) \ ,
\end{align} 
where $\Psi_4$ and $\Phi_4$ are the twist-four distribution amplitude of the proton \cite{Braun:2000kw}. ${\cal E}_g$ can be written as,
\begin{align}
{\cal E}_{ g}=2 \tilde {\cal E} +\tilde {\cal E}'~,
\end{align}
where $\tilde {\cal E}'$ is obtained from $\tilde{\cal E}$ by interchanging $y_1$ and $y_3$.  The detailed calculation yields
  \begin{align}
&\tilde{\cal E}_{\Psi }(\{x\},\{y\})=\frac{ -C_B^2  M^2_p }{12(-t)^3} (4\pi \alpha_s)^2  \nonumber\\
\times&\Bigg[x_3 K_1 
\tilde\delta[ x_1,y_1] 
\Bigl((1+\xi )^2 x_1 \bar{x}_1+2 \left(1-\xi ^2\right) y_3 \bar{x}_1
\notag \\ &+(1-\xi
  )^2 y_1 y_2\Bigr)
+
\bar x_3 \tilde K_1 
\tilde\delta[ x_3,y_3] \Bigl((1+\xi)^2 x_3 \bar{x}_3
\notag \\&
+(1-\xi)^2 y_3 \bar{y}_3\Bigr) 
+
x_3 (\tilde K_2- K_2) 
\tilde\delta[ x_2,y_2] 
\notag \\ 
&\Bigl ((1+\xi )^2 x_2 \bar{x}_2+(1-\xi)^2 y_2 \bar{y}_2
\Bigr) 
+
K_3
\tilde\delta[x_1,y_1] 
\notag \\ 
&  \Bigl
 (2 \left(1-\xi ^2\right) \bar{x}_1-(1-\xi )^2 y_1 \Bigr )
+x_3 (K_4 + K_5)
\notag \\&
 \tilde\delta[x_1,y_1]
 \Bigl (2 \left(1-\xi ^2\right) \bar{y}_1
+(1+\xi)^2 x_1 \Bigr) 
+(\tilde K_4+\tilde K_5) 
\notag \\&
\tilde\delta[ x_3,y_3] \Bigl ((1+\xi )^2 x_3 \bar{x}_3+(1-\xi )^2 y_3 \bar{y}_3 \Bigr )
\Bigg]+(\xi \leftrightarrow -\xi)~,
 \end{align}
and $\tilde {\cal E}_{\Phi}=\tilde{\cal E}_{\Psi }(1\leftrightarrow 3)$. Here we have used the following  notation to express the delta function of $x$:
 \begin{align}
     \tilde \delta[a,b]\equiv &
    \delta[x- (a-b+\xi(a+b-1))]
    \notag \\ 
 &   +    \delta[x+(a-b+\xi(a+b-1))]~.
 \label{eq:delta}
 \end{align}
The functions $K_{i}$ are the same as those defined in Eq.(\ref{eq:K}).




\end{document}